\documentclass[12pt]{article}
\usepackage{epsfig}

\newcommand{\la}{\lambda}

\begin{document}

\title{On the type of the temperature phase transition in $\phi^4$ model}
\author{M.~Bordag\thanks{Email: michael.bordag@itp.uni-leipzig.de}$~^a$, V.~Demchik\thanks{Email: vadimdi@yahoo.com}$~^b$,
 A.~Gulov\thanks{Email: alexey.gulov@gmail.com}$~^b$,
 V.~Skalozub\thanks{Email: skalozubv@daad-alumni.de}$~^b$\\~\\~
 ~$^a$ {\small \sl Institute for Theoretical Physics, University of Leipzig, Leipzig, Germany}\\
 ~$^b$ {\small \sl Dnipropetrovsk National University, Dnipropetrovsk, Ukraine}}

\maketitle

\begin{abstract}
The temperature induced phase transition is investigated in the
one-compo\-nent scalar field $\phi^4$ model on a lattice by using
Monte Carlo simulations. Using the GPGPU technology a huge amount
of data is collected that gives a possibility to determine the
Linde-Weinberg low bound on the coupling constant $\la_0$ and
investigate the type of the phase transition for a wide interval
of coupling values. It is found that for the values of $\la$ close
to this bound a weak-first-order phase transition happens. It
converts into a second order phase transition with the increase of
$\la$. A comparison with analytic calculations in continuum field
theory and lattice simulations obtained by other authors is given.
\end{abstract}


\section{Introduction}
The temperature induced phase transition in the $O(N)$ scalar
field model with a spontaneous symmetry breaking (SSB) has a long
history of investigations. It was studied either by analytic
methods of the quantum field theory or in lattice simulations (see
Refs.\cite{Zin,Berges02,Cea02} and references therein). It was
recently observed by analytic calculations within the perturbation
theory (PT) in the daisy, super daisy and some type beyond
resummations \cite{Bordag:2000tb} that a phase transition of the
first order could occur in the $O(1)$-model. However, the lack of
the expansion parameter happens near the phase transition
temperature $T \sim T_c$ for various kind resummations. So, it is
impossible to draw a reliable conclusion about the transition type
even for small values of the coupling constant $\la$. In
Ref.\cite{Baacke:2002pi} some extended kind of resummations were
used for the $O(N)$-model, and a phase transition of the second
order was determined independently of the coupling value.
Analogous results have been obtained in Monte Carlo (MC)
simulations on a lattice. As a result, nowadays the general
believe is that the phase transition is of the second order and
the PT fails in this problem. However, in the $O(N)$-models, the
results of PT calculations coincide with the lattice MC ones in
the limit of $N \to \infty$, only
\cite{Bordag:2000tb,Petropoulos:2004bt}.

Recently, a new powerful computational platform -- General Purpose
computing on Graphics Processing Units (GPGPU) technology -- has
been put in force \cite{Demchik:2009ni, Demchik:2010fd} that gives
a possibility to generate extremely large amount of MC data.
Therefore the accuracy of calculations can be essentially
increased and it becomes possible to shed light upon hidden
peculiarities and details of different processes of interest
studied by MC simulations. One of such unsolved problems is the
kind of the temperature phase transition in the $O(1)$-model for
small values of $\lambda$ and the reliability of the PT results.
This is because there are no estimates what coupling values should
be considered as small ones. As a rule, the small values are
chosen to be of the order $\la \sim 0.1 - 0.01$. However, to make
the correct choice some physical motivation is needed. One of
possible reasons is the so-called Linde-Weinberg bound on the
scalar field mass \cite{Linde:1975sw,Weinberg:1976pe}. Many years
ago these authors observed that in models with the negative mass
squared, $m^2 \le 0$, the SSB did not happen for the coupling
values below some scale, $\la \leq \la_0$. The actual value
$\la_0$ depends on the mass parameter entering the Lagrangian. So,
it is natural to consider the values of the coupling $\la \sim
\la_0$ as small ones. These values appear to be much smaller than
the values mentioned above.

In the present paper we investigate the temperature induced phase
transition in the $O(1)$-model in a wide interval of the coupling
constant $\la$ using the GPGPU technology. We obtain the
Linde-Weinberg bound $\la_0\sim 10^{-5}$. Then we compare the MC
simulations with the hot and cold starts. In a narrow interval
just above the Linde-Weinberg bound, $\la_0 \le \la ~\le
\la_1\simeq 10^{-3}$, an order parameter shows a hysteresis
behavior near the phase transition temperature. This behavior
means a phase transition of the first order. With the further
increasing of $\la$ the hysteresis behavior becomes less
pronounced and disappears at all reflecting a phase transition of
the second order.

The paper is organized as follows. In sect.2 we describe the model
and its realization on a lattice. In sect.3 a necessary
information on the MC simulations is given and the obtained
results are adduced. Sect.4 summarizes the results.

\section{The model}
In order to construct a self-consistent lattice version of the
$\phi^4$-model we start with  quantum field theory in
continuous space. The thermodynamical properties of the model are
described by the generating functional
\begin{equation}\label{gen-func}
Z=\int D\varphi~e^{-S[\varphi]},
\end{equation}
where $\varphi$ is a real scalar field, and the action is
\begin{equation}
S =\int
dx\left(\frac12\partial_\mu\varphi(x)\partial_\mu\varphi(x)-\frac12m^2\varphi(x)^2+\frac{\lambda}{4}\varphi(x)^4\right).
\end{equation}

The standard realization of generating functional in MC
simulations on a lattice assumes a space-time discretization and
the probing random values of fields in order to construct the
Boltzmann ensemble of field configurations. Then any macroscopic
observable can be measured by averaging the corresponding
microscopic quantity over this ensemble. However, the direct
lattice implementation of (\ref{gen-func}) encounters an evident
problem: the field $\varphi$ is distributed uniformly in the
infinite interval $(-\infty,\infty)$ and there is no random number
generator to simulate it. Of course, one can try to cut the
interval off, but such a procedure requires the knowledge of
characteristic scales of the model in order to separate the
unimportant tails from the interval of physically important
values. Instead, we prefer to rewrite the initial $\phi^4$ model
in continuum space-time in the form allowing a further
self-consistent lattice realization.

First we introduce one-to-one transformation $\varphi(U)$ to a new
field variable $U(x)$ defined in the finite interval $(0,1)$. Let
$U=0.5$ corresponds to $\varphi=0$ and $\varphi(U)= -
~\varphi(1-U)$. The generating functional in terms of $U$ reads
\begin{equation}
Z=\int DU\ \det\left(\frac{\partial\varphi}{\partial
U}\right)~e^{-S[\varphi(U)]}.
\end{equation}
The Jacobian can be included in the action as $\det
A=\exp(\mathrm{Tr}\log A)$. Then the new field $U$ can be easily
realized by a uniform random number generator.

The second step is the space-time discretization. For MC
simulations we introduce a hypercubic lattice with hypertorous
geometry. We use an anisotropic lattice with the spatial and
temporal spacings $a_s$ and $a_t=a_s/\zeta$, $\zeta>1$. The scalar
field is defined in the lattice sites. As a result, the generating
functional becomes
\begin{equation}
Z=\int \prod_{x} dU(x)
\exp{\left[-\left(S[\varphi(U(x))]-\sum_x\log\varphi'[U(x)]\right)\right]},
\end{equation}
where $\varphi'(U)=d\varphi/dU$ and
\begin{eqnarray}
&&S[\varphi(U(x))]= \nonumber\\
 \sum_x\frac{a_s^4}{\zeta}&& \left[
 \left(\varphi'[U(x)]\right)^2
 \frac{\partial_\mu U(x)\partial_\mu U(x)}{2}
 -\frac{m^2}{2}\varphi^2[U(x)]+\frac{\lambda}{4}\varphi^4[U(x)]\right].
\end{eqnarray}
The lattice forward derivative is defined as usually by the finite
difference operation
\begin{eqnarray}
\partial_\mu U(x)\to
\frac{U(x+a_\mu\hat{\mu})-U(x)}{a_\mu},
\end{eqnarray}
where $a_\mu$ is the lattice spacing in the $\mu$ direction,
$\hat{\mu}$ is the unit vector in the direction indicated by
$\mu$.

In the case of pure condensate field the action is determined by the
potential
\begin{eqnarray}
\tilde{V}[U]=\left[-\log\varphi'[U]+ \frac{a_s^4}{\zeta}
\left(-\frac12m^2\varphi^2[U]+\frac{\lambda}{4}\varphi^4[U]\right)\right].
\end{eqnarray}
This potential is topologically equivalent to the potential
$V(\varphi)=-m^2\varphi^2/2$ $+\lambda\varphi^4/4$. It has one
local maximum at $U=0.5$ and two symmetric global minima at $U_0$
and $1-U_0$. The spread between the values of the potential at the
local maximum and the global minima is
\begin{equation}\label{DV}
\Delta_V= \log\frac{\varphi'[U_0]}{\varphi'[0.5]}
-\frac{a_s^4}{\zeta}
\left(-\frac12m^2\varphi^2[U_0]+\frac{\lambda}{4}\varphi^4[U_0]\right).
\end{equation}

The quantities $U_0$ and $\Delta_V$ play a crucial role in MC
simulations. Being equivalent in theory, different choices of
these parameters can produce drastically different results in
actual simulations. The reason is the finite number of simulations
in an actual computer experiment. If rare but physically important
events could be missed, then the MC algorithm will not converge to
the Boltzmann ensemble of configurations. In case of $\sim 10^3$
iterations all the important probabilities have to be greater than
$10^{-3}$.

Considering the phase transition, one must guarantee that the MC
algorithm meets the field values compatible with both the phases
to choose. If $U_0\to 0.5$, then the broken phase can be missed
since the corresponding field values are extremely rare events. On
the other hand, in the limit $U_0\to 0$ (or $U_0\to 1$) the
unbroken phase is washed out. It is also important to ensure a
finite probability of the transition between those field values.
The acceptance of non-zero condensate values of the field is ruled
approximately by $\exp{(\Delta_V)}$ at each lattice site. If
$\Delta_V\gg 1$, then the unbroken phase never occurs in actual
simulations. If $\Delta_V\to 0$, then there is no broken phase. To
study the phase transition in the model, we choose the following
conditions:
\begin{equation}\label{cond-m}
U_0=0.25, \qquad \Delta_V=1.
\end{equation}
Thus, the half of generated field values will be between the
global minima of the `effective' potential, and no phase will be
accidentally missed. The probability to prefer condensate or
non-condensate values will be of order $\sim 0.5$ ensuring the
fast convergence of MC algorithm. Of course, the choice
(\ref{cond-m}) is not optimal for temperatures far away from the
critical temperature.

To satisfy two conditions (\ref{cond-m}) we use a convenient
two-parameter function,
\begin{eqnarray}\label{funct}
\varphi[U]&=&m\xi\mathrm{arctanh}\left[\eta(2U-1)+(1-\eta)(2U-1)^3\right]
\end{eqnarray}
with $\xi>0$ and $\eta>0$. The values of $\xi$ and $\eta$ have to
be found as the solution of equations $d\tilde{V}/dU|_{U=U_0}=0$
and (\ref{DV}). These equations can be written as
\begin{eqnarray}\label{syst-1}
&&\frac{2{\cal F}[U_0]-{\cal G}[U_0]}{\left({\cal F}[U_0]-{\cal G}[U_0]\right)^2}=z =\frac{\lambda\zeta}{m^4 a_s^4},\\
&&\frac{\lambda}{m^2 z}\varphi^2[U_0] ={\cal F}[U_0]-{\cal
G}[U_0],\label{syst-2}
\end{eqnarray}
where $z$ is a dimensionless parameter of the model,
\begin{eqnarray}\label{auxF}
{\cal
K}[U]&=&\frac{1-\eta(2U-1)-(1-\eta)(2U-1)^3}{1+\eta(2U-1)+(1-\eta)(2U-1)^3},\\
{\cal
F}[U]&=&
\left(\frac{{\cal K}''[U]{\cal K}[U]}{({\cal K}'[U])^2}-1\right)\log{\cal K}[U],\\
{\cal G}[U]&=& 4\left(\log\frac{-{\cal K}'[U]}{4\eta{\cal
K}[U]}-\Delta_V\right),
\end{eqnarray}
where the primes denote derivatives with respect to $U$. Eq.
(\ref{syst-1}) gives $\eta(z)$, then $\xi$ can be found from
(\ref{syst-2}). There is no physical solution for
$z<z_\mathrm{min}$. This forbidden interval corresponds to low
temperatures which cannot be reached within the chosen
parametrization. Finally the lattice action is
\begin{eqnarray}
{S}[U(x)]&=&\sum_x\sum\limits_{\mu} \left[
Y\sqrt{\frac{z}{\zeta\lambda}}
 \left(\frac{{\cal K}'[U(x)]}{{\cal K}[U(x)]}\right)^2
 \left(\frac{U(x+a_\mu\hat{\mu})-U(x)}{a_\mu/a_s}\right)^2\right]
 \nonumber\\
&+&
 \sum_x\left[-\frac{1}{4}{\cal G}[U(x)]  -Y\log^2 {\cal K}[U(x)]
 +Y^2 z\log^4 {\cal K}[U(x)] +V_0
\right],\nonumber\\
Y&=&\frac{{\cal F}[U_0]-{\cal G}[U_0]}{2\log^2{\cal K}[U_0]},\\
V_0&=& -\Delta_V -\log(2m\xi\eta).
\end{eqnarray}
The constant part of the action $V_0$ is completely unimportant
for calculations and can be omitted, since MC algorithm is based
on the difference between the actions of modified and initial
field configurations.

\begin{figure}
    \centering
    \includegraphics[bb=115 524 308 647,width=0.6\textwidth]{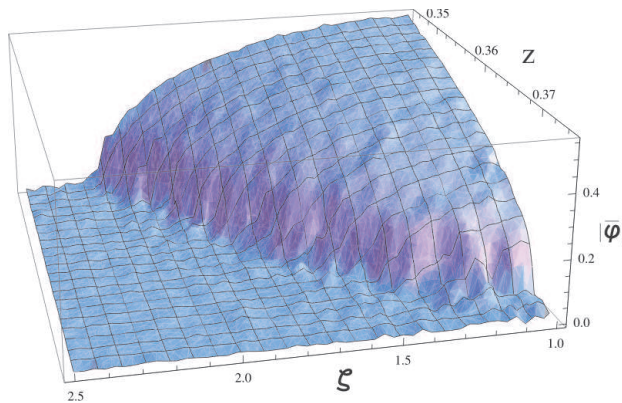}
  \caption{$|\bar\varphi|$ in the units of classical condensate
  $m/\sqrt{\lambda}$ for $\lambda=5\cdot 10^{-4}$ on lattice $16^4$}
  \label{fig:lambdaZ}
\end{figure}

By varying $\zeta$ it is possible to change $a_t$, while keeping
$a_s$ fixed. Consequently the temperature $T\sim \zeta$ can be
changed continuously at the fixed $a_s$. At low temperatures the
global discrete symmetry $\varphi\to-\varphi$ is expected to be
broken due to a non-zero field condensate. The field condensate,
$\bar{\varphi}$, can be measured directly as the average of
$\varphi(x)$ over the lattice. As an example, in
Fig.\ref{fig:lambdaZ} we plot $|\bar\varphi|$ in the units of the
classical condensate $m/\sqrt{\lambda}$ for $\lambda=5\cdot
10^{-4}$ and $16^4$ lattice. The lower values of $\zeta$ and $z$
correspond to lower temperatures. One can see the evident phase
transition with the field condensate growing with the temperature
decreasing. Since the clear positive or negative values of
$\bar\varphi$ appear in lattice configurations in the broken
phase, we conclude about the absence of domains and will use
$|\bar\varphi|$ in plots in what follows.

The field condensate is an obvious order parameter vanishing in
the high-temperature phase. We can use it to determine the type of
the phase transition. In case of the first-order transition the
overheated and supercooled states are possible. So, the MC
simulations with the hot and cold starts have to lead to different
phases near the critical temperature. Combining the MC simulations
for the hot and cold starts we will see a hysteresis plot. This
exfoliation in the vicinity of the critical temperature has to be
observed independently of the direction in the $(\zeta-z)$ plane.

We consider in details two slices of the two-dimensional function
in Fig.\ref{fig:lambdaZ} at fixed $z=0.35$ and $z=0.5$. We compute
the field condensate with the hot and cold starts for different
$\zeta$. The hysteresis-type plots will mean a first order phase
transition.

\section{Monte Carlo simulation results}

The exfoliation of the simulated data in the vicinity of the
critical temperature is a tiny effect. A large amount of
simulation data must be prepared to observe it. In this regard,
achieving the highest performance of the computational hardware is
a problem of great importance. To speed up essentially the
simulation process we apply a GPU cluster of AMD/ATI Radeon GPUs:
HD5870, HD5850, HD4870 and HD4850. The peak performance of the
cluster is up to 8 Tflops. The low-level AMD Intermediate Language
(AMD IL) is used in order to obtain the maximal performance of the
hardware. Some technical details of MC simulations on the ATI GPUs
and review of AMD Stream SDK are given in
Ref.~\cite{Demchik:2009ni} and references therein.

The MC simulations are realized at hypercubic lattices up to
$64^4$. Most of the obtained statistics come from the lattice
$16^4$. We use the pseudo-random number generator \textsf{RANLUX}
in the MC kernel, but all the key results are checked with
\textsf{RANMAR} generator \cite{Demchik:2010fd}. The lattice data
are stored with the single precision. Updating the MC
configurations are also performed with the single precision,
whereas all the averaging measurements are carried out with the
double precision to avoid the accumulation of errors.

It should be noted that for a rough estimate of the parameter
space in the early stages of the work we used a pseudorandom
number generator \textsf{XOR128} to speed up the calculations. The
condensate values obtained by this generator coincide with the
correspondent points produced with the help of \textsf{RANMAR} and
\textsf{RANLUX} generators. However, the generator \textsf{XOR128}
realized the two minima of the broken phase with an unequal
frequency. Its unexpected behavior had been discovered in the
first time, since good statistical properties of this generator
were reported in the literature on real modeling before.

The system is thermalized by passing 5000 MC iterations for every
run. For measuring we use 1024 MC configurations separated by 10
bulk updates.

\begin{center}
\begin{figure}
    \centering
    \includegraphics[bb=619 440 855 573,width=0.4\textwidth]{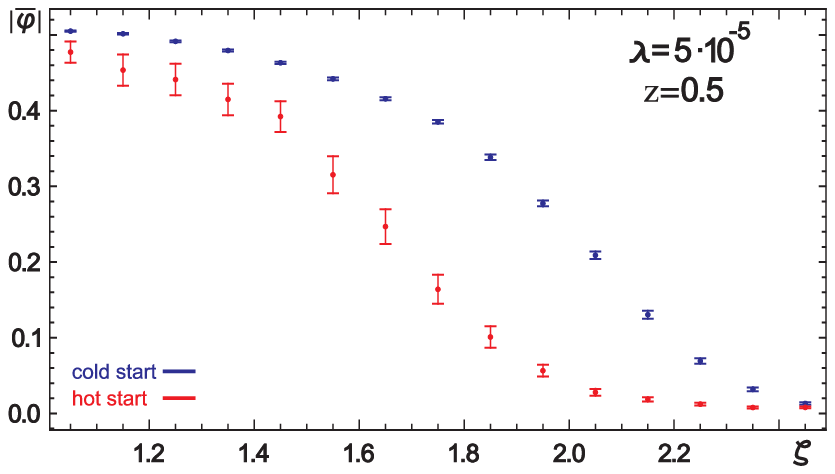}
\hskip 1cm
    \includegraphics[bb=304 617 540 749,width=0.4\textwidth]{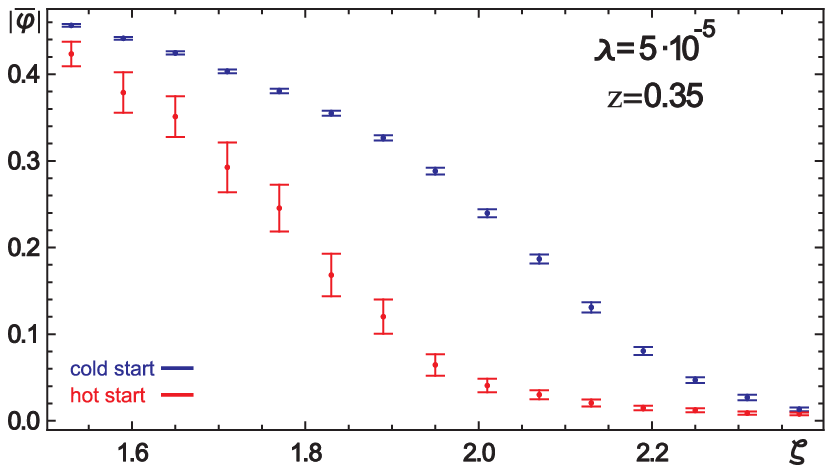}
\\\vskip 0.5cm
    \includegraphics[bb=304 440 540 573,width=0.4\textwidth]{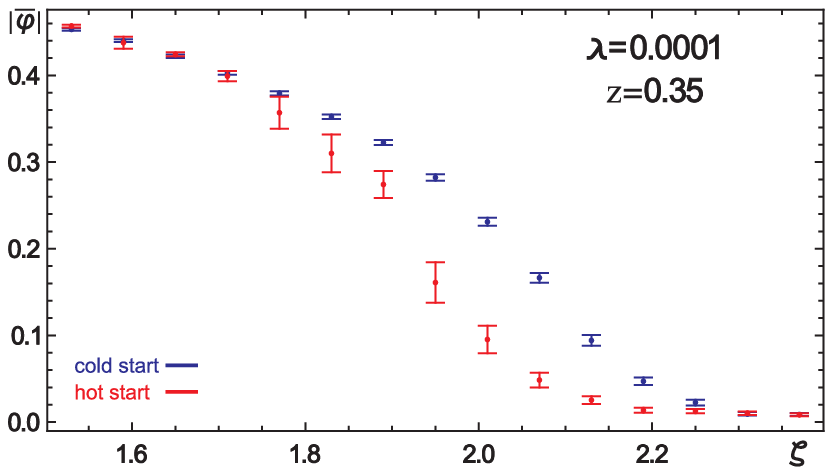}
\hskip 1cm
    \includegraphics[bb=619 617 855 750,width=0.4\textwidth]{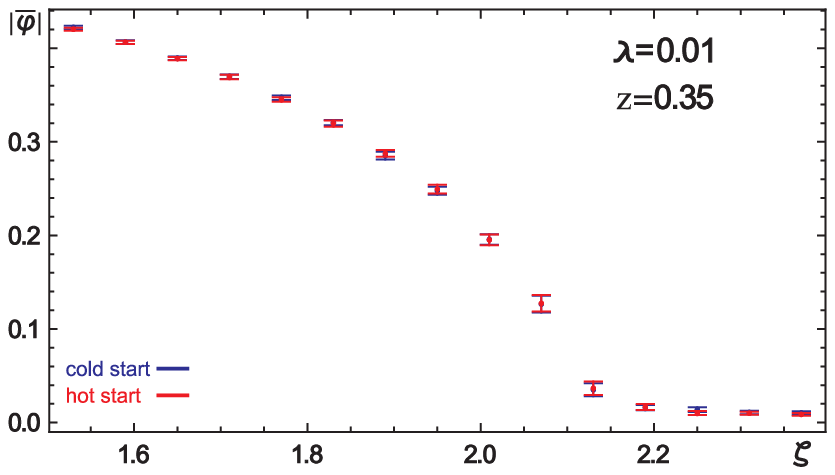}
  \caption{Temperature dependence of the absolute value of the averaged field $|\bar\varphi|$ for the lattice
$16^4$ at $z=0.35$ and $z=0.5$ for $\zeta=[1.5; 2.4]$.}
\label{fig:lambdas}
\end{figure}
\end{center}

As it was discussed in the previous section, we collect the data
for the absolute value of the averaged field $|\bar\varphi|$
representing the field condensate. The temperature dependence of
$|\bar\varphi|$ for the lattice $16^4$ at $z=0.35$ and $z=0.5$ for
$\zeta=[1.5; 2.5]$ is shown in Fig.\ref{fig:lambdas}. The whole
data set for every plot is divided into 15 bins. Different initial
conditions are marked with different colors: the hot start is
depicted in red (lower bins) and the cold start is represented in
blue (upper bins). The mean values and the 95\% confidence
intervals are shown for each bin. Every bin contains 150 simulated
points.

As it is seen from Fig.\ref{fig:lambdas}, for $\lambda=0.01$ the
temperature dependence of the field condensate is insensitive to
the start configuration chosen. Both the cold and hot starts lead
to the same behavior of the field condensate for various $\zeta$.
This means the phase transition to be of the second order, and
this result is in agreement with the common opinion on the type of
the phase transition stated in Refs. \cite{Zin, Berges02,
Baacke:2002pi}. However, therein this value of the coupling is
considered as a small one.

Then, for smaller values of $\lambda$ the overheated
configurations occur in the broken phase for the hot start, and
the supercooled states can be found for the cold start. That is,
the exfoliation of the simulated data in the vicinity of the
critical temperature is observed for different start
configurations. Such a hysteresis behavior corresponds to the
phase transition of the first order.

\begin{figure}
    \centering
    \includegraphics[bb=619 279 855 413,width=0.4\textwidth]{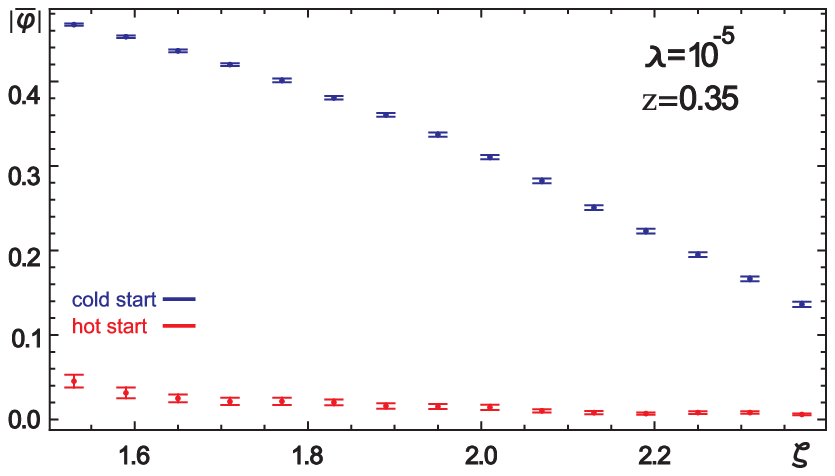}
\hskip 1cm
    \includegraphics[bb=304 279 540 413,width=0.4\textwidth]{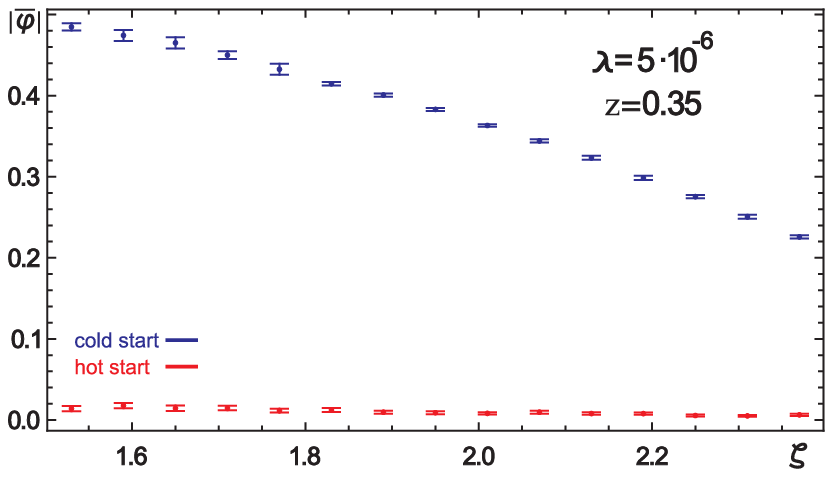}
  \caption{The Linde-Weinberg regime for the lattice $16^4$ at $\lambda=10^{-5},~ 5\cdot 10^{-6}$,
  $z=0.35$ and $\zeta=[1.5; 2.4]$.}
  \label{fig:lambdaLW}
\end{figure}

With further decreasing of $\la$ to the values of order $\la_0
\sim 10^{- 5}$ the behavior of hot- and cold-started simulations
becomes completely separated and independent of the temperature.
This effect is plotted in Fig.\ref{fig:lambdaLW}. Such type
property means that the SSB does not happen even at zero
temperature. The corresponding value $\la_0$ can be identified as
the Linde-Weinberg low bound.

Comparing the plots for two different values of $z$ with the same
set of other parameters in Fig.\ref{fig:lambdas}, one can see that
the hysteresis behavior for $z = 0.5$ is similar to the case of $z
= 0.35$. This is the common feature for all the tested values of
$\lambda$, but we adduce only one slide for brevity. The
independence of $z$ is important. It means that the observed
exfoliation of simulated points is stable within our
parametrization of the action. Thus, the introduced approach is
reliable at the transition temperature.

\section{Conclusion}
As it was discovered in MC simulations, the temperature phase
transition in the $O(1)$ $\phi^4$ model is strongly dependent on a
coupling value $\la$. There is the low bound $\la_0 \sim 10^{- 5}$
determining the range where the SSB is not realized. Close to this
value in the interval $10^{-5} \leq \la \leq 10^{- 3}$ the phase
transition is of the first order. For larger values of $\la$ the
second order phase transition happens. These observations, in
particular, may serve as a criterium for applicability of
different kind resummations in perturbation theory. In fact, we
see that the daisy and super daisy resummations give qualitatively
correct results for small values of $\la$. For larger values they
become non-adequate to the second-order nature of the phase
transition. In this case other more complicated resummation
schemes should be used.

\noindent{\bf Acknowledgements.} One of us (VD) was supported by
DFG under Grant No BO1112/17-1. He also thanks  Institute for
Theoretical Physics of Leipzig University for kind hospitality.

\end{document}